# A two-state kinetic model for the unfolding of single molecules by mechanical force

F. Ritort†, C. Bustamante‡§, and I. Tinoco, Jr.§¶∥

†Department of Physics, University of Barcelona, Diagonal 647, 08028 Barcelona, Spain; ‡Department of Physics and Molecular and Cell Biology, Howard Hughes Medical Institute, and ¶Department of Chemistry, University of California, Berkeley, CA 94720; and §Physical Biosciences Division, Lawrence Berkeley National Laboratory, Berkeley, CA 94720

**We investigate the work dissipated during the irreversible unfolding of single molecules by mechanical force, using the simplest model necessary to represent experimental data. The model consists of two levels (folded and unfolded states) separated by an intermediate barrier. We compute the probability distribution for the dissipated work and give analytical expressions for the average and variance of the distribution. To first order, the amount of dissipated work is directly proportional to the rate of application of force (the loading rate) and to the relaxation time of the molecule. The model yields estimates for parameters that characterize the unfolding kinetics under force in agreement with those obtained in recent experimental results. We obtain a general equation for the minimum number of repeated experiments needed to obtain an equilibrium free energy, to within $k_BT$, from nonequilibrium experiments by using the Jarzynski formula. The number of irreversible experiments grows exponentially with the ratio of the average dissipated work, $\overline{W_{dis}}$ to $k_BT$.**

Manipulation of individual biomolecules has led to new insights about their stability and their interactions in biologically relevant processes (1, 2). For example, force is applied to the ends of a single molecule of DNA to stretch it from a coil to an extended form, and force is applied to DNA or RNA hairpins to break base pairs and convert double-stranded regions into single strands (3–6). These "pulling experiments" are beginning to provide individual molecular trajectories amenable to treatment by simple statistical-mechanical models. In these experiments a new reaction coordinate (the end-to-end distance of the molecule under tension) can be used to describe the kinetics of the unfolding process. Recently, the pulling of RNA hairpins has shown the existence of two-state hopping transitions under force, and new kinetic effects such as hysteresis cycles, possible intermediate states, and sensitivity of specific binding sites to metal ions in solution (7, 8). Most of these experiments are done in nonequilibrium conditions where the folding-unfolding molecular relaxation time for the transition is slow compared with the time scale of the pulling experiment. In these conditions the unfolding process is irreversible. Although this nonequilibrium regime could be relevant to *in vivo* conditions where unfolding events are carried out by protein machines such as helicases and ribosomes, information about the thermodynamics parameters of the unfolding process is highly desirable. Unfortunately, current experimental limitations (thermal drift and instrumental stability) may not allow the pulling process to be slower than the molecular relaxation time. Thus, novel strategies to obtain the equilibrium parameters of the reaction from nonequilibrium realizations of the process are needed.

To this end, a remarkable identity was proposed by Jarzynski (9) that describes how equilibrium free energies $\Delta G$ can be obtained from nonequilibrium experiments by averaging the exponential of the work $W$ done on the system over the nonequilibrium trajectories, $\exp(-\Delta G/k_BT) = \overline{\exp(-W/k_BT)}$. This identity is related to a previously discovered fluctuation theorem that quantifies transient violations of the second law of thermodynamics (10, 11). A recent experimental test of this identity used the mechanical unfolding of a single RNA molecule, P5abc (8). This study showed that the Jarzynski average of successive nonequilibrium (irreversible) pulling work trajectories provides an estimate of the free energy change of the unfolding reaction within an error of order $k_BT$. Moreover, the study also provides measurements of the average dissipated work that gives additional information about the nature of the irreversible component of the process.

The purpose of the present article is 2-fold. We want to show how to obtain useful information about the kinetics of the unfolding process by studying the mean and the variance of the work probability distribution over many pulling trajectories. We also want to illustrate, in the framework of the present model, what conditions must be satisfied for the Jarzynski average to be a useful and practical method to estimate, from nonequilibrium trajectories, equilibrium free energies within an error no larger than $k_BT$.

## Model and Methods

Because our main goal is to provide simple expressions, experimentally accessible, we will analyze the simplest model that incorporates the minimum required set of parameters necessary to fit the experimental data. We model an RNA molecule as a two-level system with an intermediate barrier as depicted in Fig. 1. The two-level system corresponds to the energies of the folded and unfolded molecule, $E_f, E_u$, separated by an intermediate energy barrier of height $B$. The unfolded state has a constant length $x_m$ longer than the folded state and is thus favored by increasing force. The force necessary to unfold the hairpin in a reversible process if $\Delta E_0/k_BT \gg 1$ is approximately $F_t = \Delta E_0/x_m$, where $x_m$ is the distance between the folded and the unfolded hairpin, along the reaction coordinate. Two-level systems are good approximations for small DNA (12) or RNA (13) hairpins, known to display strong unfolding-refolding cooperativity.

At constant temperature and pressure, the differential change in Gibbs free energy, $dG$, of a system when force is applied is equal to the differential reversible work, $Fdx$. However for our two-state system with the unfolded state of constant length, $x_m$, it is convenient for us to define a Legendre transformation, $G - Fx$, so that the effect of force on the system is an effective work, $-xdF$. We use a microscopic description of the two-level system in terms of an integer variable $\sigma$, which takes the values 0 or 1 if the molecule is folded or unfolded. In terms of these variables the free energy change of the two-level system at a given force $F$ is expressed as $-Fx_m\sigma$. In a pulling experiment the molecule starts in a folded state and the force is progressively increased from zero up to a maximum force $F_m$. Different dynamical trajectories are then generated by pulling many times the same type of molecule according to the same protocol. Our main goal is to evaluate the probability distribution function for all of the possible values of work done.

---

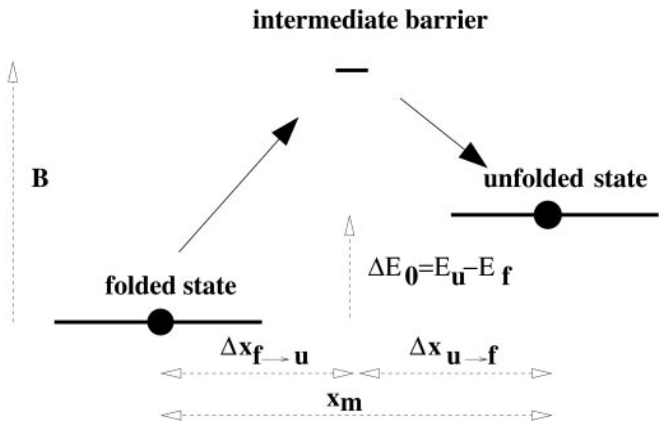

**Fig. 1.** The two-state model with an intermediate barrier. The parameters are: the energy gap $\Delta E_0$, the unfolding distance $x_m$, the intermediate barrier height $B$, and the distance $\Delta x_{f\to u}$ of the intermediate barrier to the folded state (and $\Delta x_{u\to f} = x_m - \Delta x_{f\to u}$).

For the sake of simplicity we have chosen to discretize the time in integer steps $k$ separated by an elementary time step $t_0$. Although continuous-time stochastic approaches are possible (14), our choice results in an easier procedure to carry out the path integral. The pulling protocol is described by the time evolution of the force $\{F_k; 1 \leq k \leq M\}$, where $M$ is the total number of pulling steps and $F_M = F_m$. A dynamical path is described by the sequence $\{\sigma_k; 0 \leq k \leq M\}$. To get the continuum limit we take $M \to \infty$ at the end of the calculation. The work done on the molecule** on a given path $\{\sigma\}$ is defined by (15, 16),

$$W(\{\sigma\}) = -x_m \sum_{k=0}^{M-1} \Delta F_{k+1} \sigma_{k+1}, \quad [1]$$

where $\Delta F_{k+1} = F_{k+1} - F_k$. A part of the total work $W$ is used as reversible work $W_{rev} = \Delta G$ and another part is dissipated in the form of heat $W_{dis}$. Although the second law of thermodynamics states that the average work, $\overline{W} \geq \Delta G$, or equivalently, the average dissipated work, $\overline{W_{dis}} \geq 0$,[††] for small systems nothing prevents the existence of rare dynamical paths with $W_{dis} < 0$. For example, in our two-state model, the molecule can unfold at zero or small force, thus gaining energy from the thermal bath instead of dissipating it. It is our purpose here to investigate those deviations around the average behavior and in particular those rare deviations that are responsible for the validity of the Jarzynski (9) relation $\overline{\exp(-\beta W_{dis})} = 1$ in the limit of infinite number of pulls. The probability distribution we want to evaluate, $P(W)$, is defined by,

$$P(W) = \int \mathcal{D}[\sigma] P(\sigma_0, \sigma_1, \ldots, \sigma_M) \delta(W - W(\{\sigma\})), \quad [2]$$

where $D[\sigma]$ denotes a summation over all possible dynamical paths. $P(W)$ is the probability that the total work done on the

---

**In this protocol we have chosen the force ensemble where force is the externally controlled parameter while the position $\sigma$ is the fluctuating variable. However, this is not a drawback as it can be seen that in the other ensemble where force fluctuates but position is fixed (and work is given by the expression $W(\{\sigma\}) = x_m \Sigma_{k=0}^{M-1} F_k \Delta \sigma_{k+1}$) results are equivalent.

[††]For the force ensemble it is more appropriate to refer to $\overline{W_{dis}} \geq 0$ as dissipated heat or more generally as entropy production.

molecule on mechanical unfolding is equal to $W$. To evaluate Eq. 2, we carry out the following steps:

1. Introduce a mathematical representation of the delta function.
2. Use the Bayes relation $P(\sigma_0, \sigma_1, \ldots, \sigma_M) = P_0(\sigma_0) \Pi_{j=0}^{M-1} p_j(\sigma_j, \sigma_{j+1})$ where $p_j(\sigma_j, \sigma_{j+1})$ is the conditional probability for the molecule to be in state $\sigma_{j+1}$ at time $j+1$ if it is in state $\sigma_j$ at the preceding time $j$. The initial state is $P_0(\sigma_0)$. The probabilities depend on the rates for a transition from the folded to unfolded state and vice versa. The transition rates in the presence of force (17) are given by $k_{f\to u}(F) = k_m k_0 \exp(-\beta(B - F\Delta x_{f\to u}))$, $k_{u\to f}(F) = k_m k_0 \exp(-\beta(B - \Delta E_0 + F\Delta x_{u\to f}))$ with $\beta = 1/k_B T$, where $k_0$ is an attempt frequency and $k_m$ represents the contributions from the experimental design (for instance, bead and handle fluctuations in the optical tweezers machine). The intermediate barrier is located at distances $\Delta x_{f\to u}, \Delta x_{u\to f} = x_m - \Delta x_{f\to u}$ from the folded and the unfolded states, respectively. Because the molecule stays in contact with a thermal bath the rate constants $k_{f\to u}, k_{u\to f}$ satisfy detailed balance, $k_{f\to u}(F)/k_{u\to f}(F) = \exp(-\beta(\Delta E_0 - Fx_m))$. Although our treatment is quite general, here we apply it to an unfolding pathway in which the unfolding transition under force is basically a first-order passage process (18, 19). For this reason in Fig. 1 we only represent the unfolding pathway. The conditional probabilities are: $p_j(0,1) = t_0 k_{f\to u}^j$, $p_j(1,0) = t_0 k_{u\to f}^j$, $p_j(0,0) = 1 - p_j(0,1), p_j(1,1) = 1 - p_j(1,0)$, where the time index $j$ replaces the explicit dependence of the transition rates on the force.

Elementary algebraic steps show then that this probability can be expressed as,

$$P(W) = \int_{-\infty}^{\infty} \frac{d\lambda}{2\pi} \exp(I(\lambda, W)) \quad [3]$$

with $I(\lambda, W) = i\lambda W + \log(s_1 P_0(0) + r_1 P_0(1))$ where $r_j$ and $s_j$ are the two components of a vector $\mathbf{v}_j = (r_j, s_j)$ that satisfies the (backward) recursive relation $\mathbf{v}_j = \mathbf{A}_j(\lambda) \cdot \mathbf{v}_{j+1}$ with,

$$\mathbf{A}_j(\lambda) = \begin{pmatrix} (1 - t_0 k_{u\to f}^{j-1}) \exp(i\lambda x_m \Delta F_j) & t_0 k_{u\to f}^{j-1} \\ t_0 k_{f\to u}^{j-1} \exp(i\lambda x_m \Delta F_j) & (1 - t_0 k_{f\to u}^{j-1}) \end{pmatrix} \quad [4]$$

and the initial condition $\mathbf{v}_{M+1} = (1,1)$. The recurrence in the $\mathbf{v}$s cannot be solved exactly, but most of the information about the probability distribution can be obtained for the case where the free energy change of the reaction, $\Delta G_0$, is much bigger than $k_B T$, which is the case we are mostly concerned with (typical unfolding free energies are many tens of $k_B T$). For this case, the moments of the distribution $P(W)$ can be evaluated by applying the saddle point method (method of steepest descents) and solving the resulting equations by a standard perturbative expansion in powers of $\lambda$. Although our result is general and valid for any initial state, we are interested in the case where the initial state $P_0(\sigma_0)$ is in equilibrium at temperature $T$. We skip mathematical details and quote only the final result that is obtained after taking the continuum time limit. The first two cumulants of the distribution are

$$\overline{W_{dis}} = -x_m \int_0^{F_m} dF' \int_0^{F'} dF'' \frac{\partial b(F'')}{\partial F''} g(F', F'') \quad [5]$$

$$\overline{W_{dis}^2} - \overline{W_{dis}}^2 = 2x_m^2 \int_0^{F_m} dF' \int_0^{F'} dF'' d(F'')(1 - d(F''))g(F', F''), \quad [6]$$

where $W = W_{dis} + W_{rev}$ and with the definitions

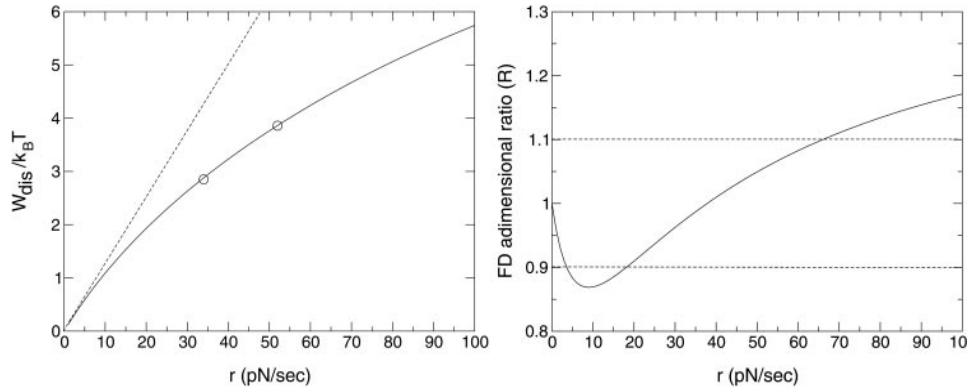

**Fig. 2.** (*Left*) Average dissipated work and comparison with reported values (circles) in P5abc from ref. 8. The dotted line is the LR approximation, Eq. **9**. (*Right*) FDR ratio *R*, Eq. **10**, for the same values showing that it is nonmonotonic and close to 1 even far from the LR regime.

$$d(F) = a(F) + \int_0^F dF' \, \frac{\partial b(F')}{\partial F'} g(F, F') \quad [7]$$

$$g(F, F') = \exp\left(-\int_{F'}^F dF'' \, \frac{k_t(F'')}{r(F'')}\right), \quad [8]$$

where $a(F) = k_{f \to u}/k_t(F)$ and $b(F) = k_{u \to f}(F)/k_t(F)$ are the unfolding and folding rate fractions and $k_t(F) = k_{f \to u}(F) + k_{u \to f}(F)$ stands for the total (folding and unfolding) rate. The term $r(F)$ is the force loading rate applied to the molecule in units of force/time. In the most general case this rate depends on the pulling speed $v$ and stiffness $k_F$ of the transducer (optical trap, cantilever, etc.) and the compliance $\partial L(F)/\partial F$ of the handles attached to the molecule via the relation: $r(F) \equiv \dot{F} = v k_F/c(F)$ with $c(F) = 1 + k_F \partial L(F)/\partial F$ being the total compliance of the transducer-handle system.[‡‡]

We can understand the physical meaning of Eq. **5** as follows. The dynamical paths in a first-time passage process can be classified in two sets: those in which the unfolding occurs at forces $F > F_t$ (i.e., those that dissipate heat into the bath $W_{\text{dis}} > 0$) and those that unfold before at $F < F_t$ (i.e., those in which heat from the bath is transformed into useful work, $W_{\text{dis}} < 0$). Each unfolding at an intermediate force $F'$ contributes an amount of $\partial b(F')/\partial F'$ to the dissipated work in **5**. It can easily be seen that all those paths with $F > F_t$ increase dissipation whereas the ones with $F < F_t$ yield a negative contribution. The memory kernel $g(F, F')$ is related to the probability that unfolding occurs between $F'$ and $F$.

Eqs. **5**–**8** can be solved exactly numerically to obtain the moments of the distribution of dissipated work for any loading rate. We can also obtain explicit equations for the limit of small loading rates. As $\beta \Delta E_0 \gg 1$ and $\Delta E_0 = \Delta G_0$ we can approximate $b(F) = -\theta(F_t - F)$, where $F_t = \Delta E/x_m$ [$\theta(x)$ is the step function $\theta(x > 0) = 1$, $\theta(x < 0) = 0$]. Note that $F_t$ is the transition force where folding and unfolding rates coincide $k_{u \to f}(F_t) = k_{f \to u}(F_t)$, therefore at that force both states are observed with the same frequency. In the slow pulling limit $r \to 0$ all integrals in **5** and **6** can be evaluated, yielding

$$\overline{W_{\text{dis}}} \simeq \rho \Delta G_0 + O(\rho^2); \qquad \rho = \frac{r}{F_t k_t(F_t)}, \quad [9]$$

where $\rho$ is a dimensionless pulling rate. The linear dependence of $\overline{W}_{\text{dis}}$ with $\rho$ defines the linear response (LR) regime. Thus, the average dissipated work for low force-loading rates is directly proportional to the pulling rate and the reversible free energy change $\Delta G_0$. It is inversely proportional to the product of the transition force $F_t$ and the total transition rate at the critical force, which is a measure of the inverse of the relaxation time of the molecular system at that force. Therefore, a plot of the average dissipated work for different pulling rates gives a measure of the fold-unfold relaxation time (at force $F_t$): $\tau_{\text{relax}}(F_t) = (k_t(F_t))^{-1}$.

### Kinetics from the Average Moments of the Dissipated Work

We now compare the prediction of this model with the main experimental results. Liphardt *et al.* (8) have measured the probability distribution of dissipated work for an RNA molecule, P5abc, in aqueous solution. The kinetics of this molecule was studied (7), although extraction of kinetic information proved difficult. At $F_t \simeq 10$ pN the molecule was found to oscillate or "hop" between the folded and unfolded configurations at an approximate rate of 10 Hz. Although P5abc does not exactly behave like a two-state state system[§§] our approach should be able to capture the kinetics of the rate-determining step [i.e., the one with smallest $k_t(F_t)$], because this is the one expected to contribute the most to the dissipated work. Actually, in ref. 8 it is shown that the slowest process occurs at $F_t \simeq 10$ pN and corresponds to a jump $x_m \simeq 10 \pm 2$ nm, giving $\Delta G_0 \simeq 25 \pm 5$ $k_B T$.[¶¶] In Fig. 2 we show the average dissipated work calculated for the best-fitting kinetic parameters. The curve was obtained by numerical evaluation of the integral, Eq. **5**, and fitting the experimental results of ref. 8 at two loading rates for the largest distance $z = 35$ nm (two points in Fig. 2). The two kinetic parameters, $k_t(F_t)$ and $\Delta x_{f \to u}$, were varied to give the best fit for a given value of $\Delta G_0$, and $F_t = 10$ pN, $T = 298$ K. A set of curves that fit the experimental data reasonably well for different values of $\Delta G_0$ (always with $\Delta G_0/k_B T \gg 1$) is found for $\Delta G_0/k_t(F_t) \simeq 1.25 \pm 0.05$ ($k_B T$ sec) and $\Delta x_{f \to u} \simeq 0.16 x_m$. Choosing $\Delta G_0 \simeq 25 \pm 5$ $k_B T$ we then find the following estimates for the kinetic parameters: $\Delta x_{f \to u} \simeq 1.6 \pm 0.4$ nm, $k_t(F_t) \simeq 20 \pm 4$ Hz (compatible with ref. 7). This set of values obtained by analyzing the average dissipated work yields a set of kinetic parameters consistent with the experimental ones. Furthermore, Fig. 2 clearly shows that the pulling experiments of ref. 8 were not done in the LR regime (shown by the dashed line).

---

[‡‡]In what follows we assume that all our estimates for the kinetic parameters include the response of the machine (bead plus handles) so they are effective kinetic parameters, our main goal being to compare our calculated results with their corresponding experimental estimates (which are also effective parameters).

[§§]Pulling curves hint at the existence of another intermediate state at forces around 14 pN (see refs. 7 and 8).

[¶¶]In ref. 8 it is estimated $\Delta G_0 = 60$ $k_B T$ for the total reversible work in the pulling experiment, but around half of the difference with the free energy of the slowest contact is due to handle contributions.

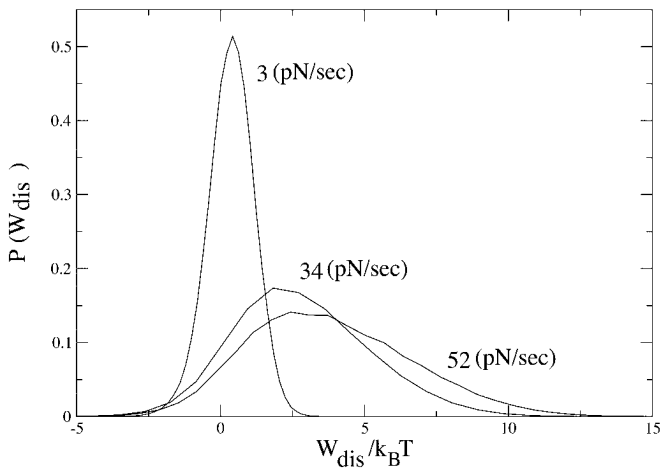

**Fig. 3.** Dissipated work probability distributions at pulling rates of 3, 34, and 52 pN/sec. They compare reasonably well with those reported in ref. 8.

Let us consider now the information contained in the variance of the distribution (8) and the fluctuation-dissipation (FD) ratio $R$, defined by the expression,

$$R = \frac{\overline{W_{dis}^2} - \overline{W_{dis}}^2}{2\overline{W_{dis}}k_BT}, \quad [10]$$

where the denominator is the magnitude of the mean-quadratic fluctuation of the dissipated work as predicted by the FD theorem (20). Thus, $R = 1$ in the LR regime and deviates as we depart from that regime. In Fig. 2 *Right* we plot the value of $R$ for different pulling rates. Comparison of this quantity with the results in ref. 8 is difficult because of the large systematic experimental errors in the SD. However, experimental results are compatible with a value of $R$ in the neighborhood of 1 and compatible with the prediction shown in Fig. 2 *Right*. Note that although $R$ is close to 1 in the whole range 1 to 100 pN/sec, its behavior is highly nonmonotonic, confirming that the pulling rates 34–52 pN/sec in ref. 8 drive the molecule far from equilibrium. This proves that, although in the LR regime $R \simeq 1$, the contrary is not necessarily true. Therefore the value of $R$ is not necessarily a good criteria to establish the LR regime. Comparison of the work distributions between the experimental results and the theory yields a remarkable similarity. Because the whole distribution (2) is difficult to compute analytically we have carried out numerical simulation pulling experiments using our model. The results are shown in Fig. 3 where we have checked that the first two cumulants of the expansion (5, 6) (shown in Fig. 2) exactly match the numerical results. Note that also the nonsymmetric shape of the distributions for 34 and 52 pN/sec is an indication that the pulling rates are far from the LR regime.

### Jarzynski Relation

What about the applicability of the Jarzynski relation (9)? By performing numerical pulling experiments we have investigated the convergence of the Jarzynski estimate $W_{dis}^{JE} = -k_BT \log(\overline{\exp(-\beta W_{dis})})$ and compared it with the two possible estimates obtained from the first and second moments of the distribution. These are the average dissipated work $\overline{W_{dis}}$ and the FD estimate, $W_{dis}^{FD} = \overline{W_{dis}}(1 - R)$, where $R$ is given in Eq. **10**. These estimates are plotted in Fig. 4) as function of the pulling coordinate (in our case the force) for the two different pulling rates used in ref. 8. Although both the FD and the Jarzynski estimates give very good estimates for the dissipated work at the maximum force, only the Jarzynski average gives a good estimate over the whole range of forces. It reproduces the free energy landscape along the entire pulling coordinate. The fact that the FD estimate works well at the maximum force is not surprising, after all $R \simeq 1$ as shown in Fig. 5). Comparing the experimental results of ref. 8, we observe that the FD estimate first increases with force, then reaches a maximum and decreases again. The value of the force corresponding to the position of the maximum dissipated work increases with the pulling rate as expected (18, 19). Interestingly, as in the experiment, the Jarzynski estimate has an error within $k_BT$ showing that the Jarzynski average can be used to recover the shape of the free energy landscape (21).

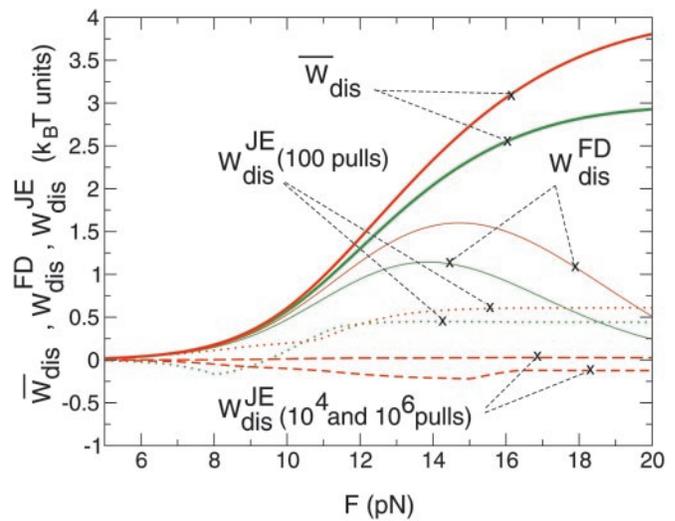

**Fig. 4.** Different estimates of dissipated work at pulling rates 34 pN/sec (green curves) and 52 pN/sec (red curves). Continuous thick lines are the average dissipated work, $\overline{W_{dis}}$, continuous lines are the FD values, $W_{dis}^{FD}$. The other noncontinuous lines correspond to the Jarzynski values, $W_{dis}^{JE}$, for $10^2$ pulls (dotted lines), $10^4$ pulls (dashed line), and $10^6$ pulls (long-dashed line). For these two last number of pulls only data for 52 pN/sec are shown.

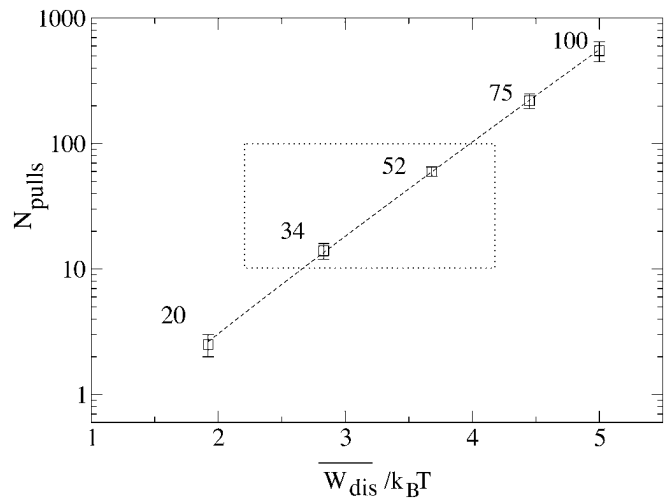

**Fig. 5.** Comparison of the number of pulls necessary to obtain an estimate for the Jarzynski average within $k_BT$ for five pulling rates in pN/sec (squares) fit to the theoretical expression, Eq. **11**. Each point corresponds to 100 sets of calculations with each set having the same number of pulls, $N_{pulls}$. The error bars show the variation among the sets. The fit to Eq. **11** yields $R_- \simeq 1.5$ in good agreement with the value obtained by analyzing the left Gaussian tail of the distributions shown in Fig. 3. The dotted box indicates the dynamical regime explored in the experiments (8).

How many pulls are needed to determine free energy within $k_BT$? The answer to this question is as follows. Consider the work distribution $P(W)$ and the variance $\Delta_-^2$ restricted to those trajectories with $W_{dis} \leq 0$ that violate the second law and define the restricted ratio $R_- = \Delta_-^2/(2k_BT\overline{W_{dis}})$. Assuming a Gaussian shape for the left tail of the distribution $P_-(W_{dis}) \sim \exp(-(W_{dis} - \overline{W_{dis}})^2/(2\Delta_-^2))$ then the Jarzynski identity $\overline{\exp(-W_{dis}/k_BT)} = 1$ holds whenever those trajectories that mostly contribute to the integral $\int_{-\infty}^{\infty} dW_{dis} P_-(W_{dis})\exp(-W_{dis}/k_BT)$ are found several times in the experiment. Because the largest contribution in this integral is peaked around the saddle point $W_{dis}^* = (1 - 2R_-)\overline{W_{dis}} < 0$ the condition to estimate the integral within $k_BT$ is given by the probability that a finite number of trajectories fall in the region $W_{dis} < W_{dis}^*$, i.e.,

$$N_{pulls} \int_{-\infty}^{W_{dis}^*} dW_{dis} P_-(W_{dis}) \sim 1.$$

This gives,

$$N_{pulls} \sim \frac{1}{\text{Erfc}\left(\sqrt{R_- \frac{\overline{W_{dis}}}{k_BT}}\right)} \sim \sqrt{R_- \frac{\overline{W_{dis}}}{k_BT}} \exp\left(R_- \frac{\overline{W_{dis}}}{k_BT}\right),$$

[11]

where $\text{Erfc}(x) = (2/\sqrt{\pi})\int_x^{\infty} \exp(-t^2)dt$ is the complementary error function. In Fig. 5 we show the number of pulls needed to estimate the free energy within $k_BT$ at five pulling rates calculated from computer simulation of our simple model. The points are fit to the theoretical expression, Eq. 11, with $R_-$ as the only parameter. We note good agreement between the computer simulations and Eq. 11. For an average dissipated work of less than $4 k_BT$, fewer than 100 experiments must be done to obtain a free energy from Jarzynski's equation. However, as the average dissipated work rises past $5 k_BT$, more than 1,000 experiments are needed.

## Conclusions

We have described a method to analyze dissipation of work during the unfolding by mechanical force of molecules that can be approximated as two-state systems. We have shown how the average dissipated work is proportional to the pulling rate in the LR regime, the constant being proportional to folding-unfolding reversible energy associated to the slowest kinetic contacts and inversely proportional to the folding-unfolding rate at the critical force. Beyond the LR regime the present approach gives estimates for the kinetic coefficients of such a two-state system from nonequilibrium experiments. This represents a practical advantage if the folding-unfolding relaxation time is so large that the pulling process cannot be done reversibly. Furthermore, because dissipated work gets contribution only from the slowest kinetic processes, the present approach provides an indirect way to recognize and isolate specific binding contacts in large molecules and estimate their free energy. Finally, we have confirmed that it is possible to use Jarzynski's identity to obtain good estimates (within $k_BT$) for equilibrium free energies by averaging over a number of pulls that asymptotically grows exponentially with the average dissipated work relative to $k_BT$, Eq. 11. This formula is general: the only system dependence is expected to enter through the constant $R_-$ that is generally of order 1.

Notice that only in the nanoscale is the Jarzynski identity useful. Beyond that scale $\overline{W_{dis}}/k_BT$ is typically much larger than 1 (e.g., of the order of the Avogadro number in a macroscopic system), so the number of trajectories needed to adequately sample the tails of the distribution is too large to be realizable. Equivalently, one could say that time scales of order of the Poincare recurrent time are needed to find some of those rare trajectories that validate the Jarzynski identity. It would be interesting to extend this analysis to the case of more complex molecules with intermediate states and more complex unfolding pathways such as multidomain proteins and RNAs.

We acknowledge useful discussions with Z. Bryant, D. Collin, S. Dumont, J. Gore, J. Liphardt, B. Onoa, and S. B. Smith. The research was supported by a North Atlantic Treaty Organization (NATO) research grant (to F.R.), Spanish Ministerio de Ciencia y Tecnologia Grant BFM2001-3525 (to F.R.), National Institutes of Health Grant GM 32543 (to C.B.), National Science Foundation Grant DBI9732140 (to C.B.), and National Institutes of Health Grant GM 10840 (to I.T.). Partial support was also provided by the David and Lucile Packard Foundation (to C.B.).